\documentclass[%
aip,
apl,
reprint,
amsmath,
amssymb,
amsfonts, 
floatfix, 
intlimits
]{revtex4-1}
\usepackage{graphicx}
\usepackage{dcolumn}
\usepackage{bm}
\usepackage{color} 
\usepackage{float}
\usepackage{lipsum}
\usepackage{cancel}
\usepackage{physics}
\usepackage{url}
\usepackage[normalem]{ulem}
\usepackage{color}
\usepackage[colorlinks = true,
	linkcolor = blue,
	urlcolor  = blue,
	citecolor = blue,
	anchorcolor = blue
	]{hyperref}
\definecolor{reviewAddColor}{RGB}{210, 42, 29}
\definecolor{reviewDelColor}{RGB}{210, 42, 29}

\begin{document}

\title{Reduced sensitivity to process, voltage and temperature variations in activated perpendicular magnetic tunnel junctions based stochastic devices}
\author{Md Golam Morshed}
\email[Author to whom correspondence should be addressed: ]{mm8by@virginia.edu}
	\affiliation{
		Department of Electrical and Computer Engineering, University of Virginia, Charlottesville, VA 22904, USA\looseness=-1}
\author{Laura Rehm}
	\affiliation{
		Center for Quantum Phenomena, Department of Physics, New York University, New York, NY 10003, USA\looseness=-1}
\author{Ankit Shukla}
	\affiliation{
		Department of Electrical and Computer Engineering, University of Illinois at Urbana-Champaign, Urbana, IL 61801, USA\looseness=-1}

\author{Yunkun Xie}
         \affiliation{San Jose, CA, 95134, USA}
\author{Samiran Ganguly}
	\affiliation{
		Department of Electrical and Computer Engineering, Virginia Commonwealth University, Richmond, VA 23284, USA\looseness=-1}
\author{Shaloo Rakheja}
	\affiliation{
		Department of Electrical and Computer Engineering, University of Illinois at Urbana-Champaign, Urbana, IL 61801, USA\looseness=-1}

\author{Andrew D. Kent}
	\affiliation{
		Center for Quantum Phenomena, Department of Physics, New York University, New York, NY 10003, USA\looseness=-1}  
\author{Avik W. Ghosh}
	\affiliation{
			Department of Electrical and Computer Engineering, University of Virginia, Charlottesville, VA 22904, USA\looseness=-1}
	\affiliation{
	    Department of Physics, University of Virginia, Charlottesville, VA 22904, USA\looseness=-1}
\date{\today}
\begin{abstract}
True random number generators (TRNGs) are fundamental building blocks for many applications, such as cryptography, Monte Carlo simulations, neuromorphic computing, and probabilistic computing. While perpendicular magnetic tunnel junctions (pMTJs) based on low-barrier magnets (LBMs) are natural sources of TRNGs, they tend to suffer from device-to-device variability, low speed, and temperature sensitivity. Instead, medium-barrier magnets (MBMs) operated with nanosecond pulses --- denoted, stochastic magnetic actuated random transducer (SMART) devices --- are potentially superior candidates for such applications. We present a systematic analysis of spin-torque-driven switching of MBM-based pMTJs ($E_b \sim~20-40~k_BT$) as a function of pulse duration ($1~\mathrm{ps}$ to $1~\mathrm{ms}$), by numerically solving their macrospin dynamics using a 1-D Fokker–Planck equation. We investigate the impact of voltage, temperature, and process variations (MTJ dimensions and material parameters) on the switching probability of the device. Our findings indicate SMART devices activated by short-duration pulses ($\lesssim 1~\mathrm{ns}$) are much less sensitive to process-voltage-temperature (PVT) variations while consuming lower energy ($\sim \mathrm{fJ}$) than the same devices operated with longer pulses. Our results show a path toward building fast, energy-efficient, and robust TRNG hardware units for solving optimization problems. 
\end{abstract}
\maketitle
True random number generators (TRNGs) are employed in many applications, including cryptography,~\cite{crypto} Monte Carlo simulations,~\cite{Monte_Carlo} neuromorphic computing,~\cite{Misra} and probabilistic~\cite{probabilistic_computing} and stochastic computing.~\cite{stochastic_computing} Conventional software algorithm-based \textcolor{black}{and CMOS-based} random number generators, \textcolor{black}{e.g., linear-feedback shift registers} do not serve the purpose of TRNG units because they produce pseudorandom bitstreams that are correlated and can be predetermined if the initial seed is known.~\cite{pseudo_rng1,pseudo_rng2,lfsr} In contrast, TRNGs utilize physical phenomena that are inherently random in nature, such as thermal noise, and radioactive decay.~\cite{thermal_noise_1,thermal_noise_2,radioactive_1,radioactive_2} Existing CMOS-based implementations of TRNGs use thermal jitter for generating true random numbers; 
however, they have large footprints and are energy-hungry.~\cite{energy_hugry_1,energy_hugry_2,energy_hugry_3}

Spintronic TRNGs provide a new opportunity in this regard.~\cite{spin_dice,review1,review2} Magnetic tunnel junctions (MTJs) constitute a fundamental building block for spintronic devices and are compatible with CMOS technology.~\cite{mtj_cmos1,mtj_cmos2} MTJs consist of two ferromagnetic layers --- a ``pinned layer" whose magnetization is fixed and a ``free layer" whose magnetization can be reoriented by a spin current --- separated by an insulator. The relative orientation between the magnetization of the pinned layer and that of the free layer gives rise to parallel (P) and anti-parallel (AP) states. The free layer of a MTJ exhibits a double potential well corresponding to two low energy states along the easy axis, separated by an energy barrier $E_b$. The magnetization state of the free layer can be switched from P to AP and vice versa by applying a current/voltage pulse, which utilizes spin-transfer torque (STT) to overcome the energy barrier.~\cite{j_sun,stt_switching,STT_andy}
Such STT-driven MTJs show a prominent stochastic switching behavior in the presence of a thermal field~\cite{Brown1963,thermal_stt} that will form the basis of our analysis.

In the past, MTJs consisting of high-barrier magnets (HBMs, $E_b > 40~k_BT$, where $k_B$ and \textit{T} are the Boltzmann constant and temperature, respectively) were frequently advocated as TRNG units; however, they suffer from high energy costs and low throughput.~\cite{spin_dice,hbm_1,hbm_2} At the opposite end of the spectrum, superparamagnetic tunnel junctions employing low-barrier magnets (LBMs, $E_b \sim k_BT$) have also been advocated as probabilistic bits. These LBMs allow the magnetization states to randomly fluctuate between P and AP, under the influence of the thermal field.~\cite{LBM_TRNG1,LBM_TRNG2,LBM_TRNG3} Although the process is very energy-efficient, LBMs suffer from slow dynamics and are rather sensitive to process and temperature variations that degrade the quality of the random bitstreams. Besides, they require near-perfect circular cross sections, and are thus hard to build in practice.~\cite{LBM_TRNG1,LBM_variability,Morshed_LBM} Stochastic magnetic actuated random transducer (SMART) devices based on perpendicular MTJs with medium-barrier magnets (MBMs, $E_b \sim 20 - 40~k_BT$) seem like a good compromise between these two extremes, for building energy-efficient and robust TRNG units.~\cite{laura_smart,ankit} However, a systematic analysis of their energy-delay-reliability-variability trade-off has not yet been undertaken to our knowledge. 

\begin{figure*}[!htbp]
\centering
\includegraphics[width=0.80\textwidth]{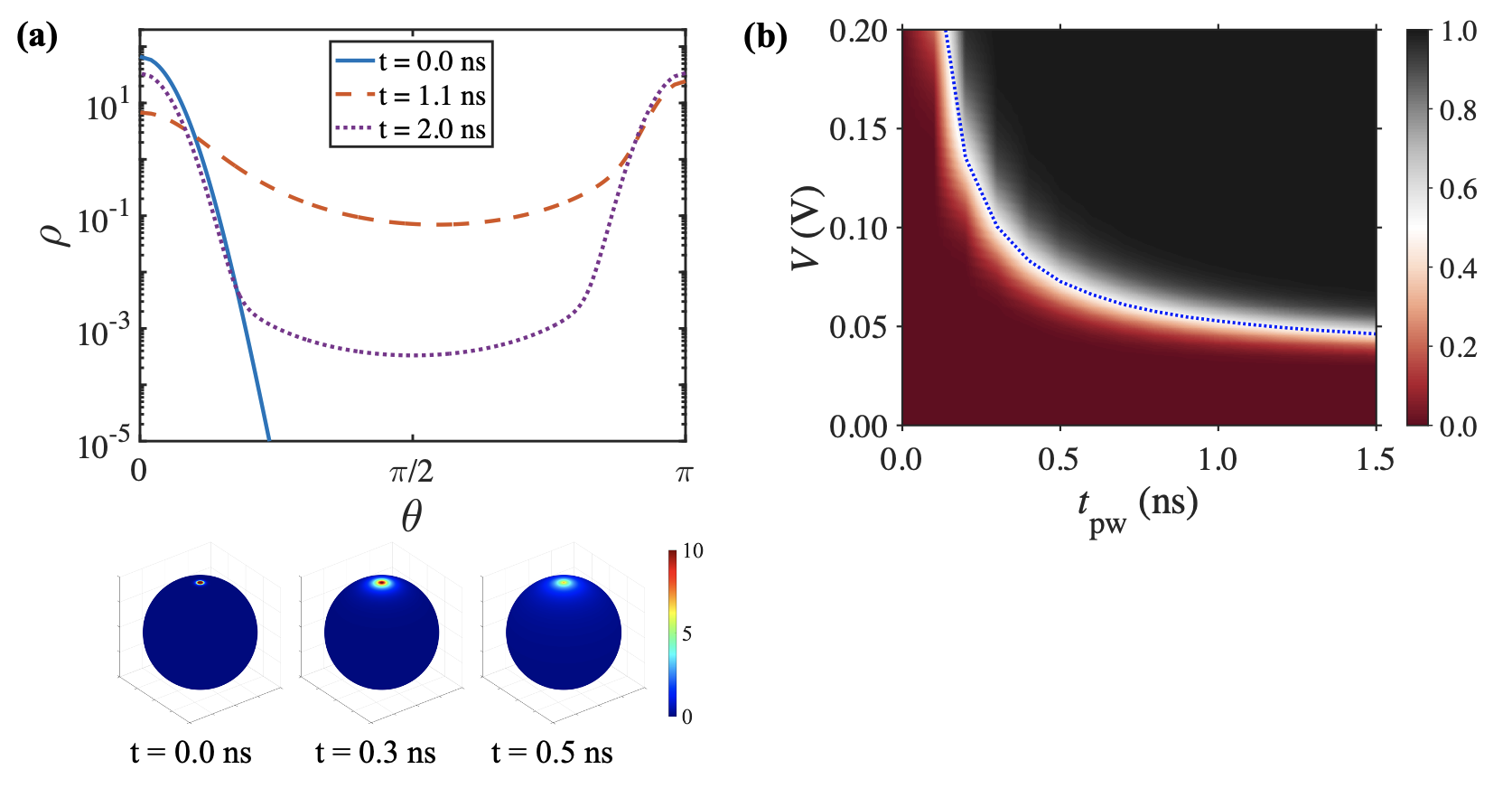}
\caption{(a) Probability density for an MBM ($\Delta \sim 35$) activated by a $1$ ns pulse (top panel). The pulse is turned ON at $t=0~\mathrm{ns}$. A bimodal distribution emerges after the pulsing, which leads to a $50\%$ switching probability. The bottom panel shows the time evolution of the probability density (colormap) during the pulsing. (b) Probability of switching (colormap) as a function of pulse amplitude and duration. The dotted blue overlaid curve represents a $50\%$ switching probability.}
    \label{fig1}
\end{figure*}

\begin{figure}[!htbp]
    \centering
     \includegraphics[width=0.90\linewidth,scale=0.5]{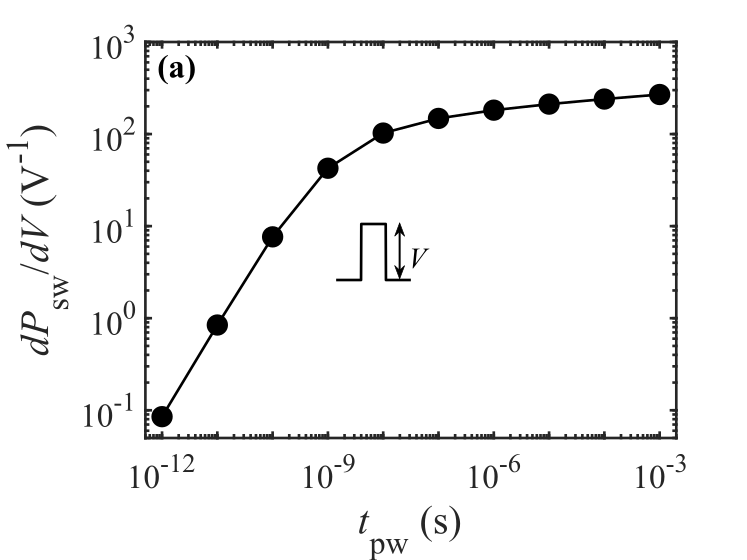}
     \includegraphics[width=0.90\linewidth,scale=0.5]{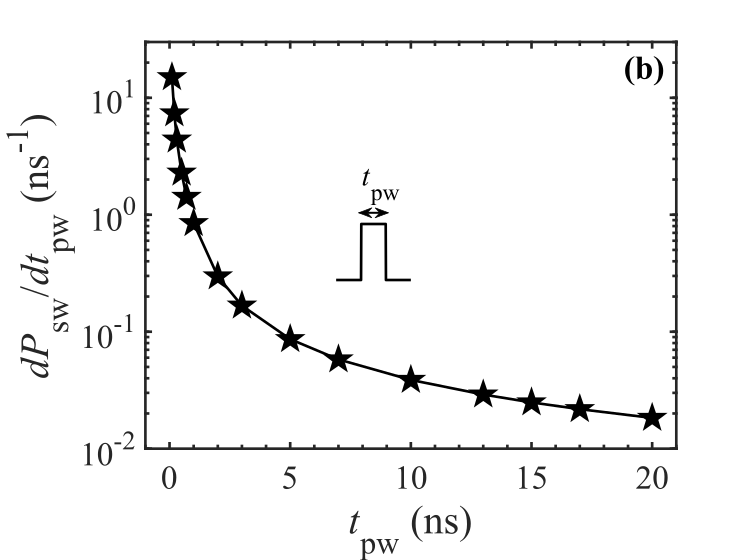}
    \caption{Variation in midpoint switching probability with respect to (a) pulse amplitude and (b) pulse duration for various pulse durations.}
    \label{fig2}
\end{figure}

\begin{figure}[!htbp]
    \centering
    \includegraphics[width=0.90\linewidth,scale=0.5]{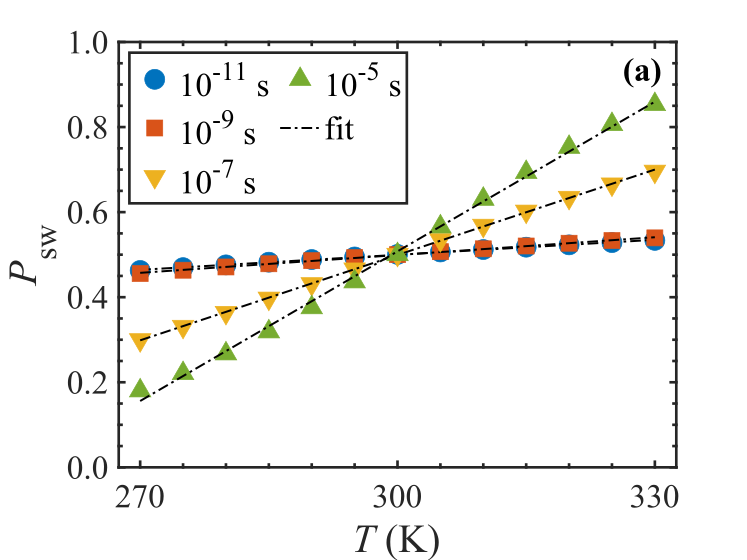}
     \includegraphics[width=0.90\linewidth,,scale=0.5]{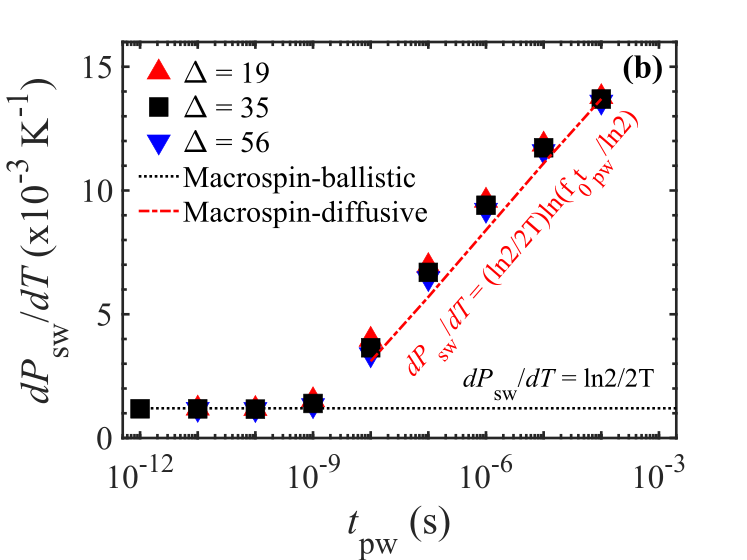}
    \caption{(a) Variation in midpoint switching probability as a function of temperature for various pulse durations. The black dash-dotted lines show the linear fit. (b) Variation in midpoint switching probability with respect to temperature for a free layer with different thermal stability factors. The black and red dash-dotted lines in (b) represent the macrospin approximation in the ballistic (short pulse) and diffusive (long pulse) limits, respectively, and the texts represent the corresponding equation.}
    \label{fig3}
\end{figure}

\begin{figure*}[!htbp]
\includegraphics[width=.32\textwidth]{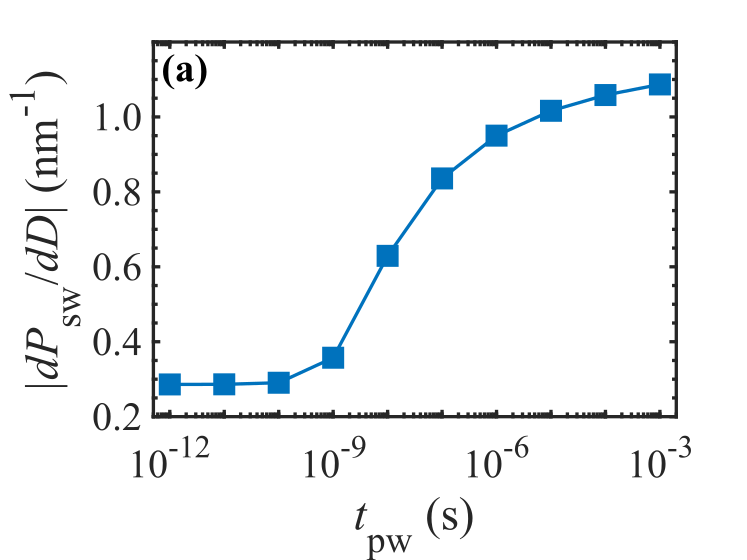}
\includegraphics[width=.32\textwidth]{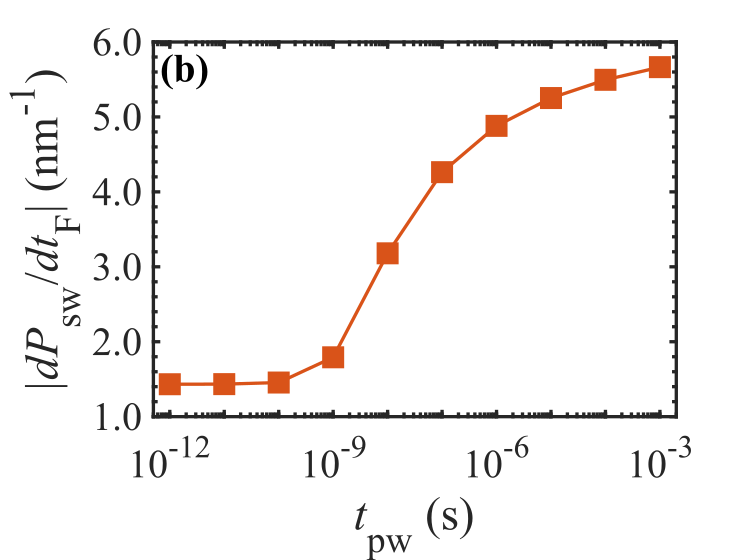}
\includegraphics[width=.32\textwidth]{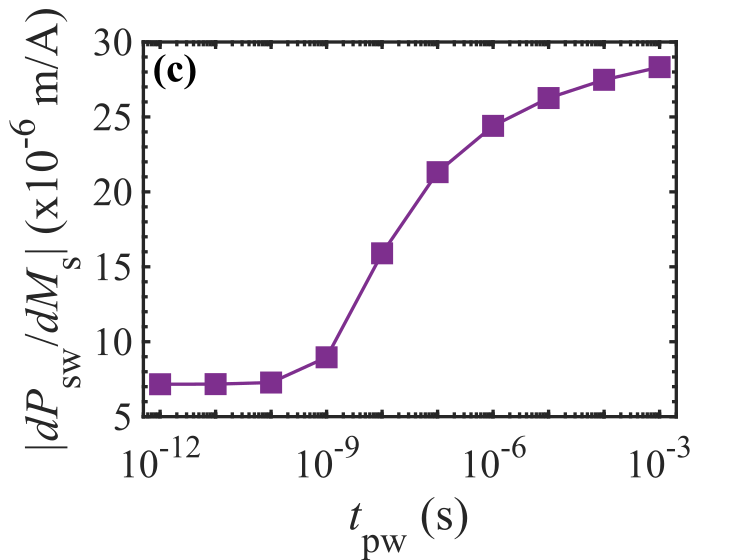}
\includegraphics[width=.32\textwidth]{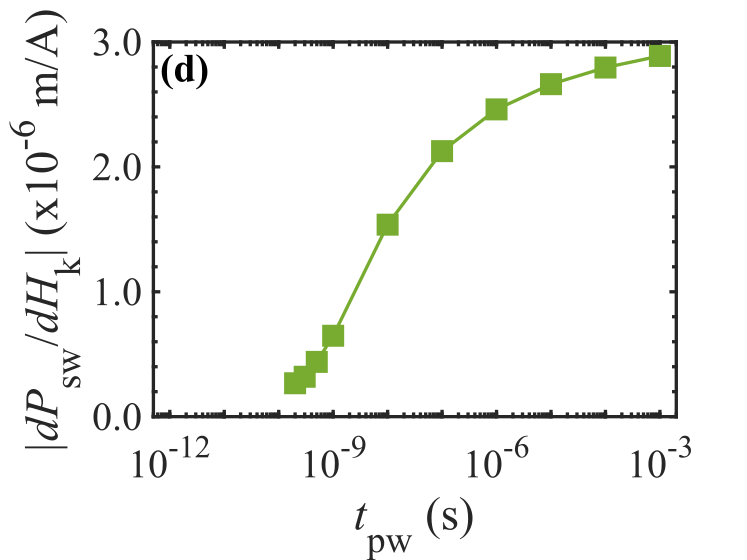}
\includegraphics[width=.32\textwidth]{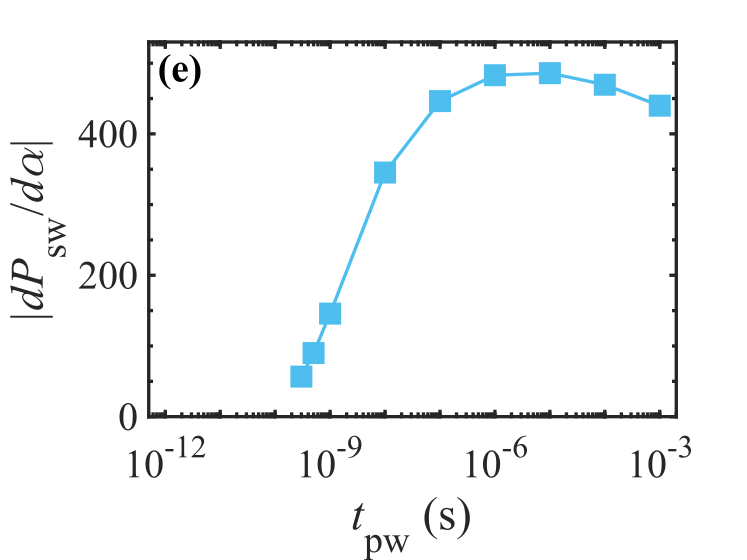}
\includegraphics[width=.32\textwidth]{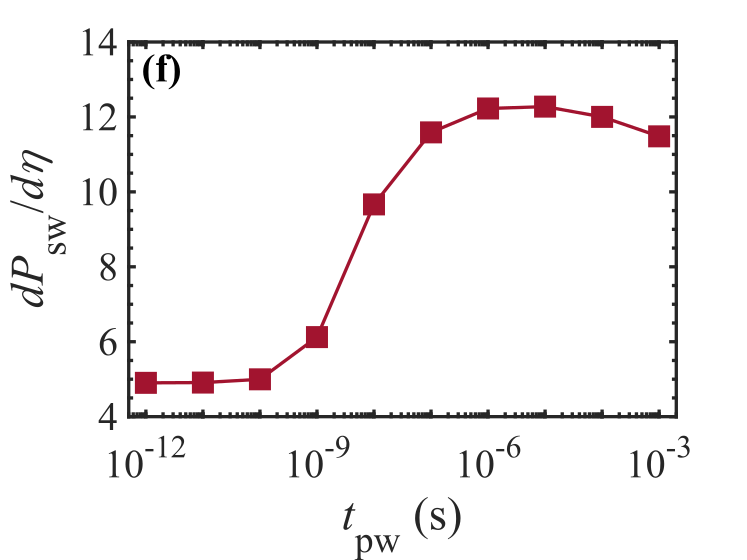}
    \caption{Variation in midpoint switching probability with respect to (a) free layer diameter, (b) free layer thickness, (c) saturation magnetization, (d) anisotropy field, (e) magnetic damping coefficient, and (f) spin polarization efficiency factor for various pulse durations. For all the variations, the change in the midpoint switching probability is lower for the short pulse limit than the longer pulse limit, leading to robust TRNG operations.}
    \label{fig4}
\end{figure*}

\begin{figure*}[!htbp]
\includegraphics[width=.32\textwidth]{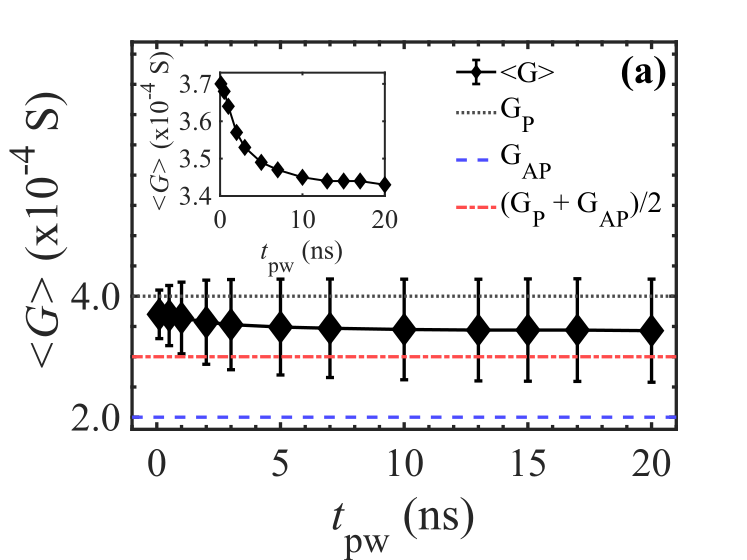}
\includegraphics[width=.32\textwidth]{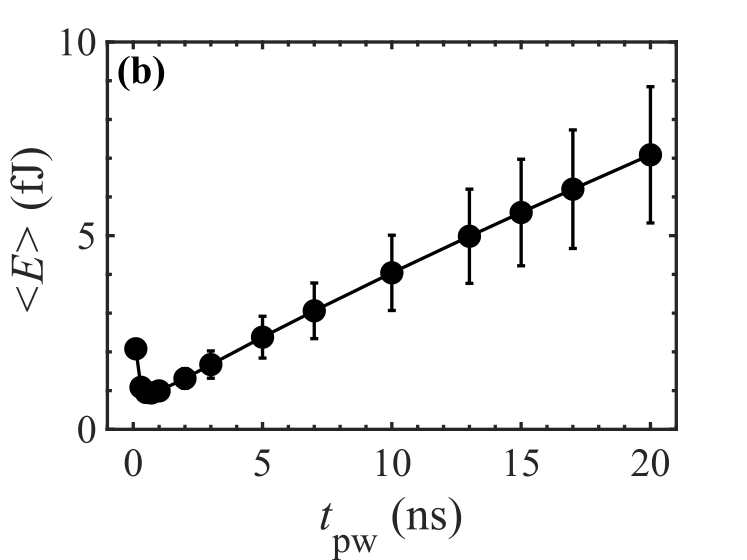}
\includegraphics[width=.32\textwidth]{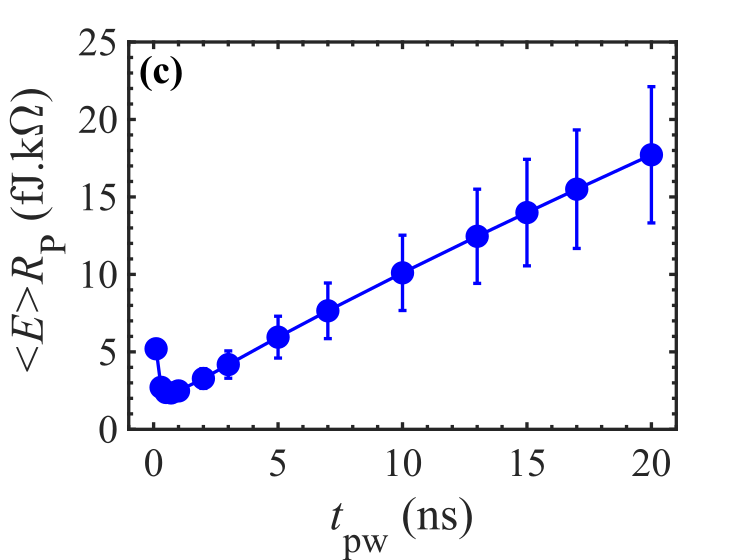}
    \caption{(a) Average junction conductance, (b) average energy dissipation, and (c) average energy-resistance product for $50\%$ switching for various pulse durations. In (a), the inset shows the zoomed view of $\langle G \rangle$. In all the figures, the error bars represent the standard deviation in our ensemble. 
    }
    \label{fig5}
\end{figure*}

In this letter, we present a comprehensive analysis of STT-driven SMART TRNGs. We numerically solve the Fokker–Planck (FP) equation to calculate the $50\%$ switching probability across a wide range of pulse durations (Fig.~\ref{fig1}) and analyze the effect of different kinds of variations on this probability. Specifically, our study investigates the influence of pulse amplitude and duration (Fig.~\ref{fig2}), temperature (Fig.~\ref{fig3}), and geometric and material parameters (Fig.~\ref{fig4}) on the $50\%$ switching probability. We find that SMART devices exhibit relatively low sensitivity to the process-voltage-temperature (PVT) variations but a greater sensitivity to pulse duration variation, especially when operated under a short-pulse limit. We estimate the energy dissipated during stochastic switching (Fig.~\ref{fig5}) and find that short pulse-activated switching consumes less energy than the same device operated with longer pulses, which suggests that SMART devices operating in the short-pulse limit can achieve robustness and energy efficiency concurrently. Our results provide a potential pathway toward the realization of fast, energy-efficient, and robust TRNG operations. 

The dynamics of magnetization is commonly described by the stochastic macrospin Landau-Lifshitz-Gilbert (LLG) equation, given by:
\begin{equation}
\begin{split}
        \frac{1+\alpha^2}{\gamma}\cdot\pdv{\vb{m}}{t} &= - \mu_0\cdot\qty(\vb{m} \cp \vb{H}_\mathrm{eff}) - \alpha \mu_0\cdot\vb{m} \cp \qty(\vb{m} \cp \vb{H}_\mathrm{eff}) \\  
        & - \frac{\hbar}{2e}\cdot\frac{\eta I}{M_s\Omega}\cdot\vb{m} \cp \qty(\vb{m} \cp \vb{m}_{p}),
\end{split}
\end{equation}
where $\vb{m} = \vb{M}/M_s$ is the normalized magnetization and $M_s$ is the saturation magnetization. \textit{I} is the applied charge current and $\vb{m}_p$ is the unit vector along the spin polarization direction. $\alpha$, $\mu_0$, $\gamma$, $\eta$, $\hbar$, $e$, and $\Omega$ are the magnetic damping coefficient, permeability of free space, gyromagnetic ratio, spin polarization efficiency factor, reduced Plank constant, elementary charge, and volume of the MTJ free layer, respectively. In the absence of an external field, the effective field $\vb{H}_\mathrm{eff}$ includes the effective anisotropy field $\vb{H}_k$, and thermal field $\vb{H}_\mathrm{th}$ ($\vb{H}_\mathrm{eff} = \vb{H}_k + \vb{H}_\mathrm{th}$). The thermal field provides a random stochastic field, which can be incorporated in a Monte Carlo solution of the above differential equation.

Alternatively and more efficiently, by solving a Fokker-Planck (FP) equation, we can quantify the statistical nature of magnetization switching under thermal fluctuations.~\cite{Brown1963,FP_butler,andy_FP,yunkun_paper} We numerically solve the 1-D differential equation form of the general FP equation:
\begin{equation}
    \pdv{\rho}{t} = -\nabla \cdot (\vb{L}\rho) + D_i \nabla^2 \rho \nonumber,
\end{equation}
where $\rho(\theta;t)$ is the probability density of the magnetization at time \textit{t},  $\theta$ being the magnetization angle to the easy axis ($z$-axis). $\vb{L}$ is the sum of all the effective torques and $D_i$ is the effective diffusive constant that accounts for the thermal fluctuations and is defined as:
\begin{equation}
    D_i = \frac{\alpha \gamma k_BT}{(1+\alpha^2)\mu_0M_s\Omega} \nonumber.
\end{equation}
The details of the numerical methods can be found in Ref.~\cite{yunkun_paper}.

The probability of switching can then be estimated from the probability density of magnetization as follows:
\begin{equation}
    P_\mathrm{sw} = \int_{\pi/2}^{\pi} \rho(\theta;t) d\theta = 1-\int_{0}^{\pi/2} \rho(\theta;t) d\theta.
\end{equation}
STT-driven magnetization switching under applied current or voltage pulses can generally be categorized into two limits --- ballistic and diffusive. Ballistic switching refers to the magnetization dynamics under short-duration pulses, while longer-duration pulses dictate the diffusive limit. In the ballistic limit, the short pulse transfers spin-angular momentum to the free layer of the MTJ, and there is little effect of thermal fluctuation during the pulsing. The switching probability in the ballistic limit for a macrospin model can be expressed as:~\cite{J_sun2,laura_smart}
\begin{equation}
    P_\mathrm{sw}^\mathrm{ballistic} = \mathrm{exp}{\left[-\frac{\pi^2\Delta}{4}\mathrm{exp}{\left\{-\left(\frac{V}{V_{c0}}-1\right)\frac{2t_\mathrm{pw}}{\tau_D}\right\}}\right]},
\label{equ3}
\end{equation}
\noindent where thermal stability factor, $\Delta = {E_b}/{k_BT} = {\mu_0H_kM_s\Omega}/{2k_BT}$ and critical voltage for switching, $V_{c0} = {2\alpha e}\mu_0H_kM_s\Omega R_P/{\eta \hbar}$, $R_P$ is the MTJ junction resistance in the P state. $\tau_D = {(1+\alpha^2)
}/{\alpha \gamma \mu_0 H_k}$ is the intrinsic time scale for the dynamics. $H_k = {2K_u}/{\mu_0 M_s}-M_s$ is the anisotropy field, where $K_u$ is uniaxial anisotropy constant. $V$ and $t_\mathrm{pw}$ are the applied pulse amplitude and duration, respectively. In our simulations, we use an MBM having $\Delta \sim 35$ unless otherwise specified. The parameters used in the simulations are listed in Table~\ref{table1}.

Figure~\ref{fig1}(a) shows the probability density of an MBM ($\Delta \sim 35$) activated by a $1~\mathrm{ns}$ pulse. {The initial quasi-equilibrium Boltzmann distribution confined near the $\theta = 0$ well ($\theta$ is the angle between the magnetization and the $z-$axis) as shown by the blue curve in the top panel of Fig.~\ref{fig1}(a)}. 
{We turn ON the pulse at $t=0~\mathrm{ns}$.} Immediately after pulsing at $t=1.1~\mathrm{ns}$, STT from the short duration pulse drives the probability density to the $\theta=\pi$ well, creating a bimodal distribution. If we relax the system for {another} nanosecond, we can clearly see the bimodal distribution leading to a $50\%$ switching probability. The bottom panel of Fig.~\ref{fig1}(a) shows the evolution of the probability density during the pulse duration. At the beginning of pulsing, the probability density is confined to the north pole of the unit Bloch sphere ($\theta = 0$) and starts spreading with time towards the south pole ($\theta = \pi$). Figure~\ref{fig1}(b) shows a colorplot of the probability of switching $P_\mathrm{sw}$ as a function of pulse amplitude $V$, and pulse duration $t_\mathrm{pw}$. As expected, the pulse amplitude required for the magnetization switching is inversely proportional to the pulse duration.~\cite{Bedau} The dotted blue overlaid curve shows the $50\%$ switching probability, which is the ideal value of the TRNG operation. We aim to operate the device near this value.

The probability of switching can be tuned through pulse amplitude, $V$, and pulse duration, $t_\mathrm{pw}$ as shown in Fig.~\ref{fig1}(b). For a specific pulse duration, we set the pulse amplitude (voltage) such that the probability of switching is $50\%$. We denote the voltage required for $50\%$ switching probability as $V_{1/2}$. From Fig.~\ref{fig1}(b), it is clear that $V_{1/2}$ will decrease as the $t_\mathrm{pw}$ increases and vice versa. However, both the $V_{1/2}$ and $t_\mathrm{pw}$ are subject to variation because, in reality, it is not feasible to apply an absolutely precise pulse amplitude and duration. We show the impact of such variations on the switching probability around the 50\% midpoint (referred to as `midpoint switching probability' hereafter) in Fig.~\ref{fig2}. {Figure~\ref{fig2}(a) shows the change in midpoint switching probability as the pulse amplitude $V$ is varied to be $\pm10\%$ of $V_{1/2}$ for various pulse durations. We find the change of midpoint switching probability with respect to $V$, $dP_\mathrm{sw}/dV$ is lower for the short-pulse limit than for the longer pulse limit. On the other hand, Fig.~\ref{fig2}(b) shows the change in midpoint switching probability to pulse duration $t_\mathrm{pw}$ for $\pm10\%$ variations in $t_\mathrm{pw}$. We get an opposite trend for the sensitivity to pulse duration. From Eq.~(\ref{equ3}), it can be shown that $dP_\mathrm{sw}/dV$ is proportional to pulse duration while $dP_\mathrm{sw}/dt_\mathrm{pw}$ is proportional to $(V/V_{c0}-1)$ around the $P_\mathrm{sw} = 50$\% value. Therefore, in the short-pulse limit, $dP_\mathrm{sw}/dV$ is lower while $dP_\mathrm{sw}/dt_\mathrm{pw}$ is higher because short pulses require larger pulse amplitudes.} 
Note that the pulse duration is kept fixed while we vary $V$. Similarly, pulse amplitude remains fixed at the corresponding $V_{1/2}$ value during pulse duration variations. Also, note that for Fig.~\ref{fig2}(b), we show data up to $20~\mathrm{ns}$ because $dP_\mathrm{sw}/dt_\mathrm{pw}$ becomes vanishingly small for longer pulses. 


Temperature plays a critical role in STT-driven MTJ switching, as it directly impacts the functionality and reliability of the device.~\cite{temp1,temp2} During the writing process, Joule heating can increase the junction temperature, which affects the device performance.~\cite{Joule_heating} The impact of thermal fluctuations mainly affects the initial magnetization distribution and the thermal stability factor ($\Delta = E_b/k_BT$). When the temperature increases from room temperature ($300~\mathrm{K}$), the thermal stability factor decreases. Therefore, for a specific $t_\mathrm{pw}$, with a fixed $V_{1/2}$, $P_\mathrm{sw}$ would be greater than $0.5$. Similarly, $P_\mathrm{sw}$ would be less than $0.5$ when temperature decreases below $300~\mathrm{K}$ as it increases the thermal stability factor. As expected, Fig.~\ref{fig3}(a) shows the linear relationship between the $P_\mathrm{sw}$ and $T$ for a $\pm 10\%$ change in the temperature from room temperature. Figure~\ref{fig3}(b) shows the $dP_\mathrm{sw}/dT$ for various pulse durations. We find that $dP_\mathrm{sw}/dT$ in the short-pulse limit is lower than for the longer-pulse limit. This higher sensitivity of $dP_\mathrm{sw}/dT$ in the diffusive limit arises from the double exponential dependence of $P_\mathrm{sw}$ on the energy barrier and temperature, 
\begin{equation} P_\mathrm{sw}^\mathrm{diffusive} = 1 - \mathrm{exp}{\{-f_0t_\mathrm{pw}\mathrm{exp}{(-E_b/k_BT)}\}},
\label{equ4}
\end{equation}
where $f_0$ is the attempt frequency. Interestingly, our FP-based result agrees well with the macrospin approximation in both ballistic and diffusive limits (Eqs.~(\ref{equ3}),~(\ref{equ4})). Specifically, in the ballistic limit, $dP_\mathrm{sw}/dT$ can be estimated as $dP^\mathrm{ballistic}_\mathrm{sw}/dT = \ln 2/2T$ while for the diffusive limit, $dP^\mathrm{diffusive}_\mathrm{sw}/dT = (\ln 2/2T) \ln(f_0t_\mathrm{pw}/\ln 2)$ around the $P_\mathrm{sw} = 50$\% value. We use $f_0 = 1~\mathrm{GHz}$, which is a commonly accepted value for magnetic materials.~\cite{Lopez-Diaz2002May} We also show the $dP_\mathrm{sw}/dT$ for the MTJ free layer with lower and higher $\Delta$ values than that of MBM ($\Delta \sim 35$). We find that $dP_\mathrm{sw}/dT$ are similar regardless of the $\Delta$ values in the ballistic limit, while there is a slight variation in the $dP_\mathrm{sw}/dT$ in the diffusive limit, where a lower $\Delta$ value results in a higher temperature sensitivity.   

We now move on to the impact of geometric and material parameter variation on the midpoint switching probability. Equation~(\ref{equ3}) suggests that for a specific pulse amplitude and duration, the probability of switching characteristic is driven by three quantities, namely $\Delta$, $V_{c0}$, and $\tau_D$. These quantities are all dependent on the geometric and material parameters of the free layer, and in practice are also susceptible to process variations during fabrication. It is thus critical to analyze the impact of parameter variation on the switching probability.~\cite{J_song}

We show the impact of variation in free layer diameter $D$ and thickness $t_F$ and material parameters $M_s$, $H_k$, $\alpha$, and $\eta$ on the midpoint switching probability in Fig.~\ref{fig4}. Among these parameters, $D$ and $t_F$ affect $\Delta$ and $V_{c0}$ proportionally through volume, while $\alpha$ and $H_k$ act like physical forces that oppose switching, affecting $\Delta$ and $V_{c0}$ proportionally and $\tau_D$ inversely. For a fixed $H_k$, the parameter $M_s$ has a similar effect on $\Delta$ and $V_{c0}$, while $\eta$ only affects  $V_{c0}$ inversely. Overall, from Fig.~\ref{fig4}, we find that for all kinds of parameter variations, the variation in the midpoint switching probability is weaker in the short-pulse limit than for longer pulses. This attribute indicates the robustness of the TRNG operation against process variations in short-pulse-activated SMART devices. While assessing parameter sensitivity, we keep $V_{1/2}$ fixed to its ideal value for a specific $t_\mathrm{pw}$ and vary only one parameter at a time. The percent variation for $D$, $t_F$, $Ms$, $H_k$, $\alpha$, and $\eta$ are $\pm 2.5\%$, $\pm 5\%$, $\pm 5\%$, $\pm 5\%$, $\pm 5\%$, $\pm 10\%$, respectively. We select these variations in ranges that each produce a linear fit with $P_\mathrm{sw}$. Also, note that for $H_k$ and $\alpha$ variations (Figs.~\ref{fig4}(d) and \ref{fig4}(e)), we exclude very low pulse duration because at such small $t_\mathrm{pw}$ values, the interplay between $\Delta$, $V_{c0}$, and $\tau_D$ changes $P_\mathrm{sw}$ in such a way that we are unable to get a linear fit to the $P_\mathrm{sw}$ data with respect to $H_k$ and $\alpha$. 

The above discussions on the $50\%$ switching probability of the short-pulse driven MBM focus on robustness. However, the other important metric during switching is energy dissipation. Moreover, from Fig.~\ref{fig2}, we can see that $dP_\mathrm{sw}/dV$ and $dP_\mathrm{sw}/dt_\mathrm{pw}$ show opposite trends as a function of pulse duration. The energy dissipation metric can set the pulse limits for the device to achieve energy efficiency and robustness simultaneously. 
The STT-driven switching in the presence of a thermal field is stochastic and the junction conductance varies in time in a stochastic way. For a constant applied voltage, we can estimate the ensemble-averaged energy dissipation $\langle E \rangle = V^2 \langle G \rangle t_\mathrm{pw}$, where $\langle G \rangle$ is the ensemble average of the junction conductance during the pulse duration. One might assume $\langle G \rangle = (G_\mathrm{P} + G_\mathrm{AP})/2$, considering half the time the magnetization is in the P state, while in the other half, it is in the AP state ($G_\mathrm{P}$ and $G_\mathrm{AP}$ are the junction conductance in the P and AP state, respectively); however, it is not guaranteed that the magnetization will spend equal time in P and AP states. An accurate way is to employ the probability density obtained by solving the FP equation, $\langle G \rangle = \frac{1}{t_\mathrm{pw}}\int_{0}^{t_\mathrm{pw}}dt\int_{0}^{\pi}d\theta \rho(\theta;t)G(\theta)$, where $G(\theta) = \frac{1}{2}(G_\mathrm{P} + G_\mathrm{AP}) + \frac{1}{2}(G_\mathrm{P} - G_\mathrm{AP})\cos{\theta}$.

Figure~\ref{fig5}(a) shows the ensemble-averaged conductance $\langle G \rangle$ for various pulse durations. For short pulse duration, $\langle G \rangle$ is close to the $G_\mathrm{P}$ value, and as the pulse duration increases, $\langle G \rangle$ value decreases because the magnetization spends more time in the AP states (see inset for zoomed view). However, $\langle G \rangle$ starts to saturate for longer pulses. We conjecture that for longer pulses, the thermal energy started taking over, which limits the evolution of probability density to the AP states regardless of the pulse duration. If we consider $\langle G \rangle = (1-k)G_\mathrm{P} + k G_\mathrm{AP}$, $k$ varies from $\sim 15 - 30\%$ as a function of pulse duration. It is noteworthy that $\langle G \rangle$ is greater than $(G_\mathrm{P} + G_\mathrm{AP})/2$ throughout the range of the pulse duration. 
From $\langle G \rangle$ we calculate the ensemble averaged energy dissipation $\langle E \rangle$  in Fig.~\ref{fig5}(b). We find a lower energy dissipation for the short-pulse limit over the longer-pulse limit. In the short-pulse limit, the energy dissipation is in the range of only a few femtojoules, which is orders of magnitude lower than the CMOS-based TRNG units (usually in the picojoules range).~\cite{cmos_energy_hungry1,cmos_energy_hungry2} {It should be noted, that we are only considering the random bit write (activation) energy here.} Finally, we quote $\langle E \rangle R_\mathrm{P}$ (in $\mathrm{fJ.k\Omega}$ unit) because it is relatively easy to vary the $R_\mathrm{P}$ in the experiment ($R_\mathrm{P} = 1/G_\mathrm{P}$ is the junction resistance in the P state). Using the tunnel magnetoresistance (TMR) relation, $\mathrm{TMR}=(G_\mathrm{P} - G_\mathrm{AP})/G_\mathrm{AP}$ in the $G(\theta)$ equation, from straightforward algebra, it can be shown that the quantity $\langle E \rangle R_\mathrm{P}$ depends only on the applied voltage, TMR, and the probability density obtained from the FP equation. In the short-pulse limit, for $100\%$ TMR, $\langle E \rangle R_\mathrm{P}$ is $\sim 5~\mathrm{fJ.k\Omega}$ or lower. Note that in Fig.~\ref{fig5}, we show data up to $20~\mathrm{ns}$ because the energy dissipation is very high for longer pulses and is unsuitable for comparison with the short-pulse limit.   

In summary, we demonstrate the suitability of MBM-based SMART devices for TRNG operations for a wide range of pulse durations. We studied the impact of various kinds of variations around the $50\%$ percent switching probability. Furthermore, we evaluate the energy consumption associated with the stochastic switching process. Our results show that the SMART devices operating in the short-pulse limit ($\lesssim 1$~ns) can achieve both robustness and energy efficiency. 
Our findings offer insights into the development of fast, energy-efficient, and reliable TRNG units for various applications.  

\begin{table}[t!]
\caption{Material parameters for MTJ free layer.}
\begin{center}
\begin{tabular}{|l|l|l|}
\hline
\textbf{Symbol}&\textbf{Definition} & \textbf{Value}\\
\hline
$D$ \ ($\mathrm{nm}$) & Diameter  & 15 \\
$t_F$ \ ($\mathrm{nm}$) & Thickness & 1.5 \\
$M_s \ \mathrm{(kA/m)}$ & Saturation magnetization & $300$ \cite{kani2016}\\
$K_u \ \mathrm{(kJ/m^3)}$ & Uniaxial anisotropy & $600$ \cite{kani2016}\\
$\alpha$ & Damping coefficient & $0.01$  \cite{kani2016}\\
$\mathrm{TMR}$ & Tunnel magnetoresistance & $100 \%$~\cite{TMR}\\
$R_\mathrm{P}~(R_\mathrm{AP}) \ \mathrm{(k \Omega)}$ & Resistance in P (AP) state&  $2.5~(5.0)$\\
$\eta$ & Spin polarization efficiency & $0.433$ \\
\hline
\end{tabular}
\label{table1}
\end{center}
\end{table}

\begin{acknowledgments}
    \vspace{-4mm}
    This work was supported in part by the NSF I/UCRC on Multi-functional Integrated System Technology (MIST) Center; IIP-1439644, IIP-1439680, IIP-1738752, IIP-1939009, IIP-1939050, and IIP-1939012. The research at NYU was supported by the DOE Office of Science (ASCR/BES) Microelectronics Co-Design project COINFLIPS and by the Office of Naval Research (ONR) under award number N00014-23-1-2771. 
    The authors at UIUC acknowledge the support of NSF through Award \# CCF-1930620 and Air Force Research Laboratory under Grant \# FA8750-21-1-0002.
    The calculations were performed using the computational resources from High-Performance Computing systems at the University of Virginia (Rivanna). 
    \end{acknowledgments}
\section*{Data Availability}
\vspace{-4mm}
The data that support the findings of this study are available from the corresponding author upon reasonable request.
\section*{REFERENCES}
\vspace{-6mm}
\bibliography{main}
\end{document}